\documentclass[showpacs,preprintnumbers,amsmath,amssymb]{revtex4}
\usepackage{graphicx,epsf}
\def\be{\begin{equation}}
\def\ee{\end{equation}}
\def\bea{\begin{eqnarray}}
\def\eea{\end{eqnarray}}
\def\A{A}
\def\Ay{{A_y}}
\def\Ayy{{A_{yy}}}
\def\R{{\cal{R}}}

\def\By{{B_{y}}}
\def\p{{\bar p}}

\def\bi{\begin{itemize}}
\def\ei{\end{itemize}}
\begin{document}
\title{
Defining perturbations on submanifolds
}
\author{Karim A. Malik, Mar\'{\i}a Rodr\'{\i}guez-Mart\'{\i}nez, 
and David Langlois}
\affiliation{ 
GRECO, Institut d'Astrophysique de Paris, C.N.R.S.,
98bis Boulevard Arago, \\
75014 Paris, France}
\date{\today}
%\maketitle

%%%%%%%%%%%%%%%%%%%%%%%%%%%%%%%
\begin{abstract}
%%%%%%%%%%%%%%%%%%%%%%%%%%%%%%%

We study the definition of perturbations in the presence
of a submanifold, like e.g.~a brane. In the standard theory of 
cosmological perturbations, one compares quantities 
at the same {\em coordinate points} in 
the non-perturbed and the perturbed manifolds, identified via 
a (non-unique) mapping between the two manifolds. 
 In the presence of a physical submanifold one needs to modify this definition
in order to evaluate perturbations of quantities at the submanifold location.
As an application, 
we compute the perturbed metric and the extrinsic curvature tensors 
at the brane position in a general gauge.

%%%%%%%%%%%%%%%%%%%%%%%%%%%%%% 
\end{abstract}
%%%%%%%%%%%%%%%%%%%%%%%%%%%%%%

\maketitle

%%%%%%%%%%%%%%%%%%%%%%%
\section{Introduction}
%%%%%%%%%%%%%%%%%%%%%%%

The theory of cosmological perturbations is a cornerstone of the study
of the early universe, since most of the accessible 
 information from this epoch is
believed to be contained in the cosmological fluctuations that are
observed in the CMB and, more indirectly, in the large scale
structures. As a consequence, an important step in the elaboration of
an early universe model is to be able to deal with the origin and 
the early evolution of cosmological
perturbations.  A recent picture
suggested (or revived) 
to describe our universe is that of a braneworld, i.e.~a
submanifold, where ordinary matter is confined, embedded in a higher
dimensional spacetime \cite{ADD,RS}. 
In a cosmological context, and for one extra
dimension, this picture has led to brane cosmology, which was shown to 
deviate from standard cosmology at high energy 
\cite{BDL,csaki,cline,SMS,BDEL}.

Soon after these progresses in homogeneous brane cosmology, several
works started to tackle the difficult problem of cosmological
perturbations in this context. One can distinguish  three main approaches
 in these various formalisms. One approach was to use
directly a doubly gauge-invariant formalism to describe the
perturbations in the bulk and in the brane \cite{muko2,Kodama,muko1,muko3}.
Another approach was to use a covariant formalism \cite{covariant,Chris,Leong}.
Finally, the most common approach has been to generalize the standard
metric based formalism 
\cite{5dpert,vandeBruck,Koyama,lmw00,david_prl,BMW1,neronov,
lmsw,deruelle,BMW2,Koyama2,grs01,cedric,riazuelo,Giudice}, 
which has been
used in standard cosmology for a long time 
\cite{Bardeen,KS,MFB}.  For detailed
reviews on braneworld perturbations see \cite{reviews}. 
In the present work, we adopt the latter approach,
which has the advantage to be of more direct access and to be more
familiar to people who have already studied the standard theory of
cosmological perturbations in four dimensions.

The specific purpose of the present work is to present how one can
describe the brane perturbations for {\it any gauge} chosen in the
{\it bulk}. Since this particular point has generated mistakes and confusion
in the literature, we believe it is worthwhile to consider this
question with some attention.  The core of the problem is that, in the
standard theory of cosmological perturbations, the perturbation for
any quantity is defined by comparing the perturbed and unperturbed
values {\it at the same coordinate point}. In brane cosmology, or in
fact in any model where one or several submanifolds have some specific
physical role, one wishes instead to define perturbations by comparing
quantities {\it at the same physical locus}.

In the next section we discuss the definition of perturbations on a
manifold and study how this definition is affected if a submanifold is
present.
In Section \ref{metric_sec} we apply these notions to a 5D metric
tensor as an example in the braneworld scenario.
In Section \ref{curv_sec} we then turn to the problem of calculating
the extrinsic curvature tensor of the brane. 
We conclude in the last section.

%%%%%%%%%%%%%%%%%%
\section{Theory}
%%%%%%%%%%%%%%%%%%%%%%

We consider the following situation: a spacetime ${\cal M}$ 
corresponding to the perturbation of a reference spacetime 
$\bar{\cal M}$ and a submanifold ${\cal S}$ in ${\cal M}$,
which can be seen as the perturbation of an unperturbed 
submanifold $\bar{\cal S}$ in $\bar{\cal M}$. We then 
wish to define meaningful (linear) perturbations
for tensorial quantities defined on the submanifold.

%%%%%%%%%%%%%%%%%%%%%%%%%%%%%%%%%%%%%%%%%%%%%%
\subsection{Standard linear perturbation theory}
\label{lintheory_sec}
%%%%%%%%%%%%%%%%%%%%%%%%%%%%%%%%%%%%%%%%%%%%%%

Let us ignore at this stage the submanifold ${\cal S}$ and let us
recall some basic principles of the standard theory of linear
perturbations in general relativity \cite{Bardeen,KS,MFB}. 
The starting point is to endow
the unperturbed manifold $\bar{\cal M}$ of dimension $N$ with a
coordinate system which we call $x^A$ (where $A=0, \dots, N-1$), which
is usually chosen according to the symmetries of $\bar{\cal M}$ so
that the explicit form of the metric is as simple as possible.

The next step is to introduce, in a geometrical language \cite{stewart},
 an identification between the
unperturbed and perturbed manifolds, i.e.~a mapping $\phi:\bar{\cal M}
\rightarrow {\cal M}$, which establishes a one-to-one correspondence
between the points of $\bar{\cal M}$ and ${\cal M}$. The
identification is not unique and thus chosen arbitrarily at first
order in the perturbations. The mapping $\phi$ naturally induces from
the coordinate system in $\bar{\cal M}$ a coordinate system in ${\cal
M}$.

%In the (coordinate based) standard theory of linear
%perturbations \cite{KS,MFB}, one defines the perturbation of 
%a tensor $T$ as
%

One then defines the perturbation of a tensor $T$ as
\be
\label{defT}
\delta T(x)= T(x)- {\bar T} (x) \,,
\ee
where $x$ stands for a given set of coordinates (which define both a
point in ${\cal M}$ and in $\bar{\cal M}$ via the mapping), 
where $T$ is the full
tensor defined in ${\cal M}$ and ${\bar T}$ is the corresponding
``unperturbed'' quantity defined in the reference spacetime $\bar{\cal
M}$.  In other words, one defines the perturbation by comparing the
values of the tensors {\it at the same coordinate point}, i.e. at the
pair of points identified by the mapping $\phi$.

As stressed above, the identification mapping is somewhat arbitrary
and one must consider slight modifications of this mapping (slight so
that one stays in the domain of validity of the perturbative approach)
to get the generic picture. A change of mapping can be interpreted as
a change of coordinates in the perturbed spacetime ${\cal M}$, which
we write as
\be
\label{coord_trans}
x^A \rightarrow \widetilde x^A=x^A+\delta x^A \,,
\ee
which means that the {\it physical point} with the old 
coordinates $x^A$ has the new coordinates $x^A+\delta x^A$.

%%%%%%%%%%%%%%%%%
\begin{figure}
\begin{center}
\includegraphics[width=70mm]{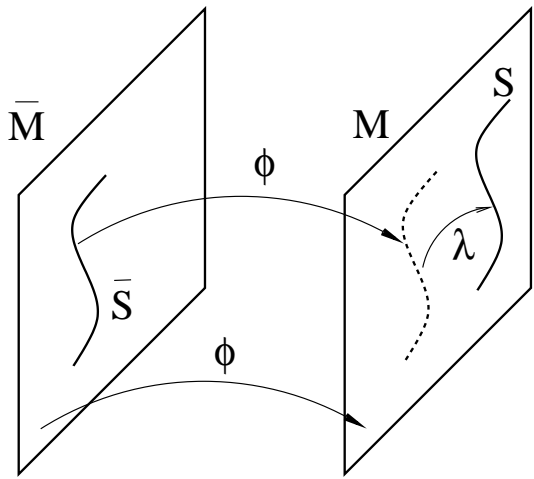} 
\caption[mapping]{
\label{mapping} 
The mapping $\phi$, which takes the background manifold, $\bar{\cal M}$, 
to the perturbed manifold ${\cal M}$, 
and the internal mapping $\lambda$, between the perturbed manifold 
and itself. To take $\bar{\cal{S}}$ to ${\cal{S}}$ we need $\phi$ and
$\lambda$.}
\end{center}
\end{figure}
%%%%%%%%%%%%%%%%

The change of any tensor under the coordinate transformation
(\ref{coord_trans}), evaluated {\it at the same old and new
coordinates} (and thus at different physical points), is given by the
Lie derivative with respect to the vector field $\delta x^A$ \cite{wald}, 
\be
\label{Ttrans}
\Delta T\equiv \tilde T(x)-T(x)=-{\cal L}_{\delta x^A} T \,.
\ee
Equation (\ref{Ttrans}) gives for the change of a covariant 2-tensor
under a transformation (\ref{coord_trans})
\be
\label{transform}
\widetilde Q_{AB}
= Q_{AB}-\pounds_{\delta x^C}  Q_{AB} \,,
\ee
or, substituting %  Eq.~(\ref{tensorlie2}) 
for the Lie derivative (see e.g. \cite{wald}), 
\be
\label{transform2}
\widetilde Q_{AB}
= Q_{AB}-\delta x^C  \partial_C Q_{AB}
-\left(\partial_A \delta x^C\right) Q_{CB}
-\left(\partial_B \delta x^C\right) Q_{AC} \,.
\ee
%

%%%%%%%%%%%%%%%%%%%%%%%%%%%%%%%%%%%%%%%%%%%
\subsection{Perturbations on a submanifold}
%%%%%%%%%%%%%%%%%%%%%%%%%%%%%%%%%%%%%%%%%%%

Let us now take into account the submanifold ${\cal S}$.  We start by
introducing the unperturbed submanifold $\bar{\cal S}$ in the
reference manifold $\bar{\cal M}$.
In general, there is no reason for the mapping $\phi$ to leave the
submanifold invariant, i.e.~we have in general $\phi(\bar{\cal S})\neq
{\cal S}$, as shown in Fig.~\ref{mapping}.
Of course, one can always choose a special subclass of mappings which do
leave invariant this submanifold but this corresponds to a restricted
choice of coordinate systems.  Note, that the Gaussian normal (GN)
coordinate system, defined below in Section \ref{GNsec}, belongs to this
subclass.

If one wishes to define perturbations for quantities existing only at
the geometrical locus of the submanifold, one therefore cannot use the
above definition, Eq.~(\ref{defT}). A necessary 
%intermediate
additional step is
to define another mapping $\lambda$ between the image of $\bar{\cal
S}$ and the real ${\cal S}$, i.e. $\lambda: \phi(\bar{\cal
S})\rightarrow {\cal S}$.  In the coordinate system defined by $\phi$,
this can be written in the form $\hat x^A=x^A+\epsilon^A$,
i.e.~$\lambda$ maps a point of $\phi(\bar{\cal S})$ with coordinates
$x^A$ onto a point of ${\cal S}$ with coordinates $x^A+\epsilon^A$.

One can now define a meaningful perturbation for tensorial quantities 
defined at the submanifold location, 
\be
\delta_{\cal S}T(x)=T(\lambda(\phi(\p)))- {\bar T}(\p) \,,
\ee
where $x$ stands for the coordinates of $\p$, which is a point of
$\bar{\cal S}$.  When the quantity $T$ is in fact defined everywhere
(but one wants to use the perturbation of its value at the location of
the submanifold), the defining expression above can be decomposed into
\be
\delta_{\cal S}T(x)=T(\lambda(\phi(\p)))- T(\phi(\p))+T(\phi(\p)) 
- {\bar T}(\p) \,, 
\ee
where one can recognize in the last two terms the usual definition of
the linear perturbation. Reintroducing coordinates, this reads
\bea
\label{Spert}
\delta_{\cal S} T(x)
&=&T(x+\epsilon)-T(x)+\delta T(x) \nonumber \\
&=&\epsilon^A\partial_A T+\delta T(x) \,,
\eea
where we have used the Taylor expansion of $T(x+\epsilon)$.  Equation
(\ref{Spert}) is an adequate definition of the perturbation of a
tensorial quantity in the presence of a submanifold, if we impose the
restriction that the perturbation has to be at the physical locus of
the submanifold.

%%%%%%%%%%%%%%%%%%%%%%%%%%%%%%%%%%%%%%%%%%%%%%%%%%%%%
\section{The metric of a braneworld}
\label{metric_sec}
%%%%%%%%%%%%%%%%%%%%%%%%%%%%%%%%%%%%%%%%%%%%%%%%%%%%%

We now apply the results found in the previous
section to the metric tensor of a braneworld.
After briefly reviewing how the perturbed metric tensor
changes under a first order coordinate transformation
we show how the definition of the perturbation of the metric tensor
changes in the presence of a particular submanifold, the brane.

%%%%%%%%%%%%%%%%%%%%%%%%%%%%%%%%%%%%%%%%%%%%%%%%%%%%%%%%
\subsection{Standard perturbed metric tensor and change 
under a gauge-transformation}
%%%%%%%%%%%%%%%%%%%%%%%%%%%%%%%%%%%%%%%%%%%%%%%%%%%%%%%%

We study a 5D metric with maximally symmetric flat 3-subspace (see for
example \cite{BMW2,riazuelo}) and include only scalar perturbations,
that is perturbations that transform like scalars on spatial
3-sections.

The background part of the metric tensor is
given by
\be
\label{backmetric}
ds^2=\bar g_{AB}dx^Adx^B
= -n^2 dt^2+a^2\delta_{ij}dx^idx^j+b^2dy^2\,,
\ee
where the metric factors $n$, $a$, and $b$ are functions 
of coordinate time $t$ and extra dimension $y$.
The metric tensor, including linear perturbations, is 
\begin{equation}
\label{pertmetric} 
g_{AB}\equiv \bar g_{AB}+\delta g_{AB}
= \left(
\begin{array}{ccc}
-n^2(1+2\A) & a^2 B_{,i} & n\Ay \\ 
a^2 B_{,j} & a^2\left[(1+2\R)\delta_{ij}+2E_{,ij}\right] & a^2 B_{y,i} \\
n\Ay & a^2 B_{y,i} & b^2(1+2\Ayy)
\end{array}
\right) \, .
\end{equation}
where the scalar metric 
potentials $A$, $B$, $\R$, $E$, $B_y$, $A_y$ and $A_{yy}$,
are functions of the coordinates $x^A=[t,x^i,y]$.

To find the change of the metric tensor under the first order
coordinate transformation defined by Eq.~(\ref{coord_trans}),
we apply Eq.~(\ref{transform2}) to the metric tensor $g_{AB}$, 
given in Eq.~(\ref{pertmetric}). This gives the perturbed 
metric tensor in the new coordinate system
\be
\widetilde \delta g_{AB}= \delta g_{AB}-
\delta x^C  \partial_C \bar g_{AB}
-\left(\partial_A \delta x^C\right) \bar g_{CB}
-\left(\partial_B \delta x^C\right) \bar g_{AC} \,,
\ee
where 
$\delta x^A = \left[\delta t, \delta x_{,}^{~i},\delta y\right]$.
Since we are only working to linear order, the Lie derivative with
respect to the first order vector $\delta x^A$ only acts on the
background part of the metric, $\bar g_{AB}$.

The transformation behavior of the scalar metric perturbations is
therefore
\bea
\label{coordtrans}
\widetilde \A   &=& \A -\dot{\delta t}-\frac{\dot n}{n}\delta t
-\frac{n'}{n}\delta y    \,, %\\ \nonumber
\qquad \widetilde B   = B + \frac{n^2}{a^2}\delta t 
-\dot{\delta x}  \,, \\ \nonumber
\widetilde \R   &=& \R -\frac{\dot a}{a} \delta t-\frac{a'}{a} \delta y   
\,, %\\ \nonumber
\qquad \widetilde  E  =  E -\delta x   \,, \\ \nonumber
\widetilde \Ay   &=& \Ay +n\delta t' - \frac{b^2}{n} \dot{\delta y}   
\,, %\\ \nonumber
\qquad \widetilde \By  = \By -\delta x' -\frac{b^2}{a^2}\delta y   \,, \\ 
\nonumber
\widetilde \Ayy   &=& \Ayy -\frac{\dot b}{b} \delta t -\frac{b'}{b}\delta y
-\delta y'    \,. 
\eea
%

%%%%%%%%%%%%%%%%%%%%%%%%%%%%%%%%%%%%%%%%%%%%
\subsection{Perturbed metric tensor in the 
presence of a submanifold}
%%%%%%%%%%%%%%%%%%%%%%%%%%%%%%%%%%%%%%%%%%%%

The above definition of the perturbed metric, Eq.~(\ref{pertmetric}),
does not depend on the presence of a submanifold and is defined
everywhere in the bulk.  However, our spacetime contains a brane and
we are interested in computing the perturbed metric on this {\it
physical} hypersurface, i.e.  we need to evaluate it on the brane.

In our model the homogeneous brane is fixed at $x^4 = const$, but
after being perturbed, it is in general displaced from this
location. We would like to describe the change in the metric at the
brane position as a result of this displacement.  
This displacement is embodied in the relation 
\be
\hat x^A=x^A+\epsilon^A \,,
\ee
where the $\epsilon^A$ are associated with 
 the mapping $\lambda$ introduced earlier.
In the general case the $\epsilon^A$ are decomposed into 
degrees of freedom tangential and orthogonal to the brane
according to (see e.g. \cite{deruelle})
\be
\label{defepsilonA}
\epsilon^A\equiv \zeta^{\mu}e^A_{\mu} + \zeta n^A  \,,
\ee
where $n^A$ is the normal vector to the brane, 
$e^A_{\mu}$ is a basis of vectors tangential to the brane, 
and $\mu = 0,...,3$.

From Eq.~(\ref{Spert}) it follows that the perturbed metric
at the position of the brane is given by
\be
\label{defcalSmetric}
\delta_{\cal S} g_{AB}
=\epsilon^C\partial_C g_{AB} + \delta g_{AB} \,.
\ee

\begin{figure}
\begin{center}
\includegraphics[width=100mm]{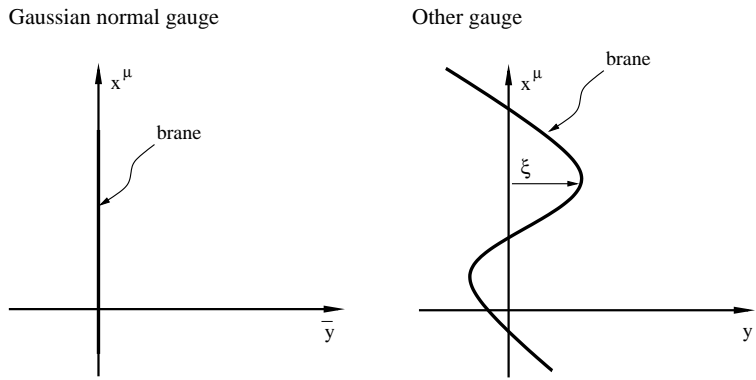} \\
\caption[gn_pic]{\label{gn_pic} 
The brane in Gaussian normal (left) and a different coordinate
system.}
\end{center}
\end{figure}

In our model the only relevant degree of freedom of $\epsilon^A$ is
the 4-component, the perturbation of the brane position in the
extra-dimension, which we denote by $\epsilon^4\equiv \xi(t,x^i)$, see
Fig.~\ref{gn_pic}. The other degrees of freedom $\zeta^{\mu}$, tangential
to the brane, correspond to mappings that leave the brane invariant, and 
 we are therefore allowed to choose  $\zeta^{\mu}\equiv 0$.

The perturbed metric in the presence of the brane is
thus given by
\be
\label{deltaSgmunu}
\delta_{\cal S} g_{AB}= \delta g_{AB}+ \zeta \,n^A \partial_A \bar g_{AB}
= \delta g_{AB}+\xi\,\partial_4 \bar g_{AB} \,.
\ee
Note, that since we only work to linear order in the perturbations,
and the perturbation $\xi$ is first order, the additional part in the
perturbed metric tensor, $\xi\,\partial_4 \bar g_{AB}$, only includes
derivatives of the background metric.

The metric tensor at the position of the brane 
is therefore given by 
\be
\label{totmetric}
g^{\cal S}_{AB}= \bar g_{AB}+\delta_{\cal S} g_{AB} \,,
\ee
where $\bar g_{AB}$ is the background metric.

%%%%%%%%%%%%%%%%%%%%%%%%%%%%%%%%%%%%%%%%%%%%%%%%%%%
\section{Extrinsic curvature}
\label{curv_sec}
%%%%%%%%%%%%%%%%%%%%%%%%%%%%%%%%%%%%%%%%%%%%%%%%%%%

In this section we calculate the extrinsic curvature tensor 
in the presence of a submanifold.
The extrinsic curvature $K_{AB}$ describes the local bending of the
brane along the extra dimension. 
In the braneworld scenario this bending is determined by the local
matter distribution on the brane through the junction conditions
\cite{israel}.

The extrinsic curvature tensor is defined as (see e.g. \cite{wald}),
\be
\label{Kmunu}
K_{AB} = h_A^{C}\,\left({}^{(5)}\nabla_C n_B\right)\,,
\ee
where $n^B$ is the unit vector normal to the brane,
${}^{(5)}\nabla_A$ is the 5D covariant derivative 
and $h_{AB}$ is the projection tensor, defined as 
$h_{AB}\equiv g_{AB}-n_A n_B $. 

We shall study the case of a static brane, i.e.~a brane that 
is not moving with respect to the background coordinate system. 
The unit vector $n^A$ is space-like and thus subject to the constraint
\be
\label{n_Aconstraint}
n_A n^A=1\,.
\ee

In the following section
we calculate the extrinsic curvature tensor  
in an arbitrary gauge up to first order in the perturbations.
As a check, we then calculate the extrinsic curvature tensor
in the Gaussian normal (GN) gauge and transform the expressions
found in this gauge to an arbitrary gauge.

%%%%%%%%%%%%%%%%%%%%%%%%%%%%%%%%%%%%%%%%%%%%%%%%%%%%%%%
\subsection{Calculating $K_{AB}$ in an arbitrary gauge}
%%%%%%%%%%%%%%%%%%%%%%%%%%%%%%%%%%%%%%%%%%%%%%%%%%%%%%%

%We will now calculate the extrinsic curvature tensor in an arbitrary
%gauge without making use of the GN gauge.
%
%Without loss of generality we will assume that we can choose
%a coordinate system in the background
%in which the brane is at rest and located at $\bar y=0$.

The constraint Eq.~(\ref{n_Aconstraint}) together with background
metric Eq.~(\ref{backmetric}) give the unit normal vector to the brane
at zeroth order as
\be
\label{backn_A}
\bar n^A=\left[0,{\bf 0},b^{-1} \right] \,.
\ee 
As pointed out above, we consider a brane that is non-moving and hence 
the time component of $\bar n_A$ is zero.

The perturbed normal vector at the brane position can be computed by
using the perturbed version of Eq.~(\ref{n_Aconstraint}),
\be
2 \bar n_A\,\delta_{\cal S} n^A 
+ \bar n^A \bar n^B \delta_{\cal S} g_{AB} = 0 \,.
\ee
It is important to stress that one has to use $\delta_{\cal S}g_{AB}$
in the above equation and not $\delta g_{AB}$ defined in
Eq.~(\ref{pertmetric}), since one is perturbing
Eq.~(\ref{n_Aconstraint}) defined \emph{on the brane}. We therefore
get for the normal vector to the brane at the position of the brane up
to first order
\be
\label{defn_A}
n_A=
\bar n_A + \delta_{\cal S} n_A =
b\left[-\dot\xi,-\xi_{,i},1+\Ayy+\frac{b'}{b}\xi \right] \,.
\ee

Substituting the background metric Eq.~(\ref{backmetric}) and the
background normal vector Eq.~(\ref{backn_A}) into the definition of
the extrinsic curvature tensor, Eq.(\ref{Kmunu}), we get for the
components of the extrinsic curvature tensor at zeroth order
\be
\label{backKmunu}
\bar K_{00} = -\frac{n^2}{b} \frac{n'}{n}\,, \qquad
\bar K_{ij} = \frac{a^2}{b} \frac{a'}{a} \delta_{ij}  \,. 
\ee
To get the extrinsic curvature tensor up to first order at the
position of the brane, $K_{AB}=\bar K_{AB} +\delta_{\cal S} K_{AB}$,
we have to use perturbed quantities defined on the brane, i.e.~the
perturbed metric $g^{\cal S}_{AB}$, defined in
Eq.~(\ref{totmetric}), and the perturbed normal vector
(\ref{defn_A}). Substituting into Eq.~(\ref{Kmunu}) then gives for the
components of the perturbed part, $\delta_{\cal S} K_{AB}$,
\bea
\label{Kmunucomp}
\delta_{\cal S} K_{00} &=& -\frac{n^2}{b}\left\{
%\frac{n'}{n}+
A'+2\frac{n'}{n}A -\frac{n'}{n} \Ayy +\frac{1}{n}\dot\Ay
+\frac{b^2}{n^2}\left[\ddot\xi
+\left(2\frac{\dot b}{b}-\frac{\dot n}{n}\right)\dot\xi\right]
+\left[
\frac{n''}{n}-\frac{n'}{n}\frac{b'}{b}+\frac{n'^2}{n^2}
\right]\xi
\right\}\,, \\ \nonumber
\delta_{\cal S} K_{0i} &=& \left\{ \frac{1}{2}\frac{a^2}{b}
\left[B'-\dot\By+2\frac{a'}{a}B-\frac{n}{a^2}\Ay\right]
-\dot\xi+\left(\frac{\dot a}{a}-\frac{\dot b}{b}\right)\xi
\right\}_{,i} \,, \\ \nonumber
\delta_{\cal S} K_{04} &=& 
\frac{1}{b}\frac{n'}{n}\left(n\Ay+b^2\dot\xi\right)
\,, \\ \nonumber
\delta_{\cal S}K_{ij} &=& \frac{a^2}{b}\left\{
%\frac{a'}{a}+
\R'-\frac{a'}{a}\Ayy
+\frac{1}{n^2}\frac{\dot a}{a}\left(n \Ay+b^2\dot\xi\right)
+\left(\frac{a''}{a}+\frac{a'^2}{a^2}-\frac{a'b'}{ab}\right)\xi
\right\}\,
\delta^i_j \\ \nonumber
&\qquad& 
+\frac{a^2}{b}\left\{E'-\By+2\frac{a'}{a}E-\frac{b^2}{a^2}\xi
\right\}_{,ij} \,, \\ \nonumber
\delta_{\cal S}K_{i4} &=& 
\frac{1}{b}\frac{a'}{a}\left(a^2\By+b^2\xi
\right)_{,i} \,, 
\qquad \delta_{\cal S}K_{44} = 0 \,. \\ \nonumber
\eea
%
%[Checked: 3/2/2003 KAM]

%%%%%%%%%%%%%%%%%%%%%%%%%%%%%%%%%%%%%%%%%%%%%%%%%%%%
\subsection{Starting from Gaussian normal gauge}
\label{GNsec}
%%%%%%%%%%%%%%%%%%%%%%%%%%%%%%%%%%%%%%%%%%%%%%%%%%%%

In the following section we outline the calculation of the extrinsic
curvature tensor in the Gaussian normal (GN) gauge. To get the components
of the extrinsic curvature tensor in an arbitrary gauge, we then have to
transform these results 
%from the GN gauge 
to an arbitrary gauge which we
shall do in the following section \ref{expansion_sec}.

%%%%%%%%%%%%%%%%%%%%%%%%%%%%%%%%%%%%%%%%%%%%%%%%%%%%
\subsubsection{Calculating the extrinsic curvature 
tensor in a Gaussian normal gauge}
%%%%%%%%%%%%%%%%%%%%%%%%%%%%%%%%%%%%%%%%%%%%%%%%%%%%

We begin by computing the extrinsic curvature tensor in a slightly
more general gauge than the GN gauge, so that we can still work
directly with the background metric defined in Eq.~(\ref{backmetric}),
denoting quantities in this gauge by a ``hat''.

In our nearly GN gauge 
the metric has the particularly simple form
\be
\widehat g_{44}=b^2 \,, \qquad \widehat g_{4\mu} = 0 \,,
\ee
which is in terms of the scalar metric perturbations 
\be
\label{defGN}
\widehat\Ay =0 \,, 
\qquad 	 \widehat\Ayy=0\,, 
\qquad   \widehat\By=0\,.
\ee
The brane is located at $\widehat y= 0$, i.e.~there is 
no perturbation in the brane position
\be
\widehat\xi=0 \,.
\ee
Note that to go from this gauge to the GN, one only needs to put $b =1$.

The normal vector to the brane is in this gauge
\be
\widehat n^A=[0,{\bf 0}, b^{-1}]\,,
\ee
and the definition of the extrinsic curvature tensor Eq.~(\ref{Kmunu})
simplifies to
\bea
\widehat K_{AB} &=& \frac{1}{2 b}\frac{\partial}{\partial \hat y} 
\widehat{g}_{AB} \qquad \mbox{for} \qquad A ,\, B \neq  4 \, ,\\
\widehat K_{A4} &=& \widehat K_{4B} = 0 \,. \nonumber
\eea
The extrinsic curvature is therefore
in components
\bea
\label{KGN}
\widehat K_{00} &=& -\frac{n^2}{b}\left[ 
\frac{n'}{n}+2\frac{n'}{n}\widehat\A+\widehat\A'\right] \,, \\
\widehat K_{0i} &=& \frac{a^2}{b}\left[ 
\frac{a'}{a}\widehat B+\frac{1}{2}\widehat B'\right]_{,i} \,, 
\nonumber \\
\widehat K_{ij} &=& \frac{a^2}{b}\left[ 
\left(\frac{a'}{a}+2\frac{a'}{a}\widehat\R+\widehat\R' \right)\delta_{ij}
+\widehat E'_{,ij}+2\frac{a'}{a}\widehat E_{,ij} \right] \,. \nonumber
\eea
Again the components in the GN gauge are obtained by putting $b=1$.

%%%%%%%%%%%%%%%%%
%%%%%%%%%%%%%%%%

%%%%%%%%%%%%%%%%%%%%%%%%%%%%%%%%%%%%%%%%%%%%%%%%%%%%%%%
\subsubsection{Transforming the extrinsic curvature to an 
arbitrary gauge}
\label{expansion_sec}
%%%%%%%%%%%%%%%%%%%%%%%%%%%%%%%%%%%%%%%%%%%%%%%%%%%%%%%

We can now transform the extrinsic curvature tensor in the hat gauge,
Eq.~(\ref{KGN}) to an arbitrary gauge by using the transformation
rules of the tensor components themselves, Eq.~(\ref{transform2}), and
by the transformation rules of the metric potentials,
Eq.~(\ref{coordtrans}).

Under a first order coordinate transformation the extrinsic curvature
tensor changes according to Eq.~(\ref{transform2}),
\be
\widetilde K_{AB}= \hat K_{AB}-\delta x^C  \partial_C \bar K_{AB}
-\left(\partial_A \delta x^C\right) \bar K_{CB}
-\left(\partial_B \delta x^C\right) \bar K_{AC} \,.
\ee
The components therefore change as
\bea
\label{transKmunu}
\widetilde K_{00}&=& \hat K_{00} -  \partial_t \bar K_{00} \, \delta t
-\bar K'_{00}\, \delta y -2\dot{\delta t}\, \bar K_{00} \,, \\
\widetilde K_{0i}&=& \hat K_{0i} -\delta t_{,i} \, \bar K_{00}
-\dot{\delta x}_,^{~k} \,\bar K_{ki} \,, \nonumber \\
%
%\widetilde K_{04}&=& \hat K_{04} -\delta t' \,\bar K_{00} \,, 
%\nonumber \\
%
\widetilde K_{ij}&=& \hat K_{ij}-\partial_t \bar K_{ij}\,\delta t
-\bar K'_{00} \,\delta y
-2 \delta x^k_{,i} \,\bar K_{kj} \,. \nonumber
\eea

{}From the definition of the hat gauge Eq.~(\ref{defGN}) and the
transformation behavior of the metric perturbations
Eq.~(\ref{coordtrans}) it follows that the hat gauge only restricts
the $y$-derivatives of the coordinate transformation $\delta x^A$,
i.e.~the degrees of freedom orthogonal to the brane, but leaves the
gauge transformations on the brane, that is tangential to it,
arbitrary \cite{BMW2}.
We can solve Eqs.~(\ref{defGN}) and (\ref{coordtrans}) for
$\delta t'$, $\delta x'$, and $\delta y'$ and get
\bea
\label{deltax'}
\delta x' &=& \By-\frac{b^2}{a^2}\delta y \,, \\ 
\delta t' &=& -\frac{1}{n}\Ay+\frac{b^2}{n^2}\dot{\delta y} \,, 
\nonumber \\
\delta y' &=& \Ayy-\frac{\dot b}{b}\delta t -\frac{b'}{b}\delta y \,. 
\nonumber
\eea

We now substitute the expression for
the extrinsic curvature tensor in the nearly GN gauge, 
Eq.~(\ref{KGN}),
into Eq.~(\ref{transKmunu}), 
use the expressions for the change of the scalar
metric potentials under a coordinate transformation,
Eq.~(\ref{coordtrans}), together with Eq.~(\ref{deltax'}) and the
expression for the background part of the extrinsic curvature tensor
Eq.~(\ref{backKmunu}), and get after a straight forward but tiresome
calculation the extrinsic curvature in an arbitrary gauge.

But we are not completely there, yet. As pointed out in Section
\ref{lintheory_sec}, a gauge transformation leaves the coordinate
point unchanged. Hence the extrinsic curvature tensor
%Eq.~(\ref{Kmunuexpanded}) 
is still at the coordinate position of the brane 
in the nearly GN gauge at $y=0$
and we therefore have to shift it to the physical position of 
the perturbed brane at $y=\xi$.
We can do this by using a Taylor expansion of $K_{AB}$
around $y=0$, and get
\be
\label{kab_taylor}
K_{AB}\left.\right|_{y=\xi}
=K_{AB}\left.\right|_{y=0}+\epsilon^C  K_{AB,C} \,,
\ee
where, see Eq.~(\ref{defepsilonA}) above, $\epsilon^A=[0,\bf{0},\xi]$ 
is the shift between the coordinate systems in which the 
brane is at $y=0$ and $y=\xi$, respectively.\\

The final piece of information that we need for the complete extrinsic
curvature tensor in an arbitrary gauge is the relation between $\delta
y$ and $\xi$, since the expression we have derived so far still
contains $\delta y$ and its time derivatives.  The relation is readily
found by calculating the change of the normal vector defined above in
Eq.~(\ref{defn_A}) under an infinitesimal coordinate transformation
$\delta x^A$, which is given by $\widetilde{n_A} = n_A -
\pounds_{\delta x^B} n_A$ \cite{wald}.
We therefore find for the transformation behavior of $\xi$ under an
infinitesimal coordinate transformation
\be
\widetilde{\xi}=\xi+  \delta y  \,,
\ee
and since in the nearly GN gauge $\widehat \xi =0$, we get 
\be
\delta y=- \xi \,.
\ee

Hence we get the same expression for the components of the perturbed
curvature tensor in an arbitrary gauge, that is Eq.~(\ref{Kmunucomp}),
by transforming from the nearly GN gauge as by direct calculation, as
expected.

%%%%%%%%%%%%%%%%%%%%%%%%
\section{Conclusions}
%%%%%%%%%%%%%%%%%%%%%%%%

In this paper we have studied the definition of first order
perturbations in the presence of a submanifold.  
Perturbations in standard cosmological perturbation theory can be
defined via  a mapping $\phi$ of tensorial quantities between a background
manifold $\bar{\cal M}$ and a perturbed manifold ${\cal M}$.
The introduction of a submanifold, or hypersurface, ${\cal S}$ 
into the spacetime does not present a problem in itself, since it doesn't 
affect the definition of the perturbations. 
A problem might arise, if we require the perturbations 
to be restricted to the submanifold, 
since in general $\phi(\bar{\cal S})\neq {\cal S}$. 
This problem can be 
alleviated, either by choosing a particular coordinate system, 
e.g.~a Gaussian Normal (GN) one,
which enforces the correct mapping between the submanifold
in the background and its perturbed image, or by leaving the
coordinate system unrestricted, but then using a second mapping 
$\lambda$ which takes the image of the submanifold, $\phi(\bar{\cal S})$ 
to the correct position of the submanifold, 
i.e.~$\lambda[\phi(\bar{\cal S})] ={\cal S}$.

%The problem of defining perturbations on a submanifold, e.g.~on a
%brane, does not arise in the GN coordiante system or gauge, since one
%restricts the perturbations to be on the brane by definition. But by
%using the GN gauge from the outset of the calculation, one looses the
%freedom to adopt a particularly suitable or simple gauge.

The usage of a non-GN gauge requires care in defining perturbations
at the physical position of the submanifold, as we have demonstrated. 
Although the problem of defining perturbations on a submanifold does not 
arise if we work in a GN coordinate system or gauge from the outset,
as already pointed out, one looses in this case the
freedom to adopt a particularly suitable gauge for the problem 
or a gauge that simplifies the calculations, by having ``used up''
some of the gauge freedom. Of course it is always possible to start in the
GN coordinate system and then transform to a different gauge, but this 
in itself is non-trivial, as we have shown as well.

As an example, we have calculated the perturbed metric tensor 
for a braneworld scenario at the position of the brane. 
Using this result we then  
calculated the perturbed extrinsic curvature tensor at the brane
position. We have also shown how to calculate the extrinsic curvature 
tensor in the GN gauge and its transformation into an arbitrary gauge.

In Ref.~\cite{muko1} Mukohyama uses a doubly gauge-invariant 
formalism to investigate the perturbed junction conditions
in braneworld cosmology. In his formalism the gauge transformations
on the submanifold, or brane, are allowed to differ from 
the gauge transformations in the bulk spacetime. This additional 
freedom is not required for most applications and makes the formalism
difficult to use. If we limit the gauge transformations 
in the bulk and on the brane to be identical, the doubly gauge-invariant 
formalism and our approach give the same results.

Although the examples given in this paper are concerned with the
definitions of perturbations on the brane in a five-dimensional bulk,
it has some connections with the question of the hypersurface matching
in standard 4D \cite{4Dmatching} or the question of the matching of
perturbations across a cosmological bounce \cite{bounce}. In these
cases, one finds a physical hypersurface and although this
hypersurface is timelike, instead of spacelike as the brane, our
approach could be certainly useful.

%%%%%%%%%%%%%%%%%%%%%%%
\acknowledgments
%%%%%%%%%%%%%%%%%%%%%%%

The authors would like to thank Helen Bridgman, David Matravers, Alain
Riazuelo, Filippo Vernizzi and David Wands for useful discussions. KM
is supported by a Marie Curie Fellowship under the contract number
\emph{HPMF-CT-2000-00981}.  Algebraic computations of tensor
components were performed using the GRTensorII package for Maple.

%%%%%%%%%%%%%%%%%%%%%%%%%%%%

%%%%%%%%%%%%%%%%%%%%%%%%%%

%%%%%%%%%%%%%%%

\begin{thebibliography}{}
%%%%%%%%%%%%%%%%%%%%%%%%%%%%

\bibitem{ADD}
N.~Arkani-Hamed, S.~Dimopoulos and G.~R.~Dvali,
%``The hierarchy problem and new dimensions at a millimeter,''
Phys.\ Lett.\ B {\bf 429}, 263 (1998)
[arXiv:hep-ph/9803315].

\bibitem{RS}
L.~Randall and R.~Sundrum,
%``An alternative to compactification,''
Phys.\ Rev.\ Lett.\  {\bf 83} (1999) 4690
[arXiv:hep-th/9906064].

\bibitem{BDL}
P.~Binetruy, C.~Deffayet and D.~Langlois,
%``Non-conventional cosmology from a brane-universe,''
Nucl.\ Phys.\ B {\bf 565} (2000) 269
[arXiv:hep-th/9905012].

\bibitem{csaki}
C.~Csaki, M.~Graesser, C.~Kolda and J.~Terning,
%``Cosmology of one extra dimension with localized gravity,''
Phys.\ Lett.\ B {\bf 462}, 34 (1999)
[arXiv:hep-ph/9906513].

\bibitem{cline}
J.~M.~Cline, C.~Grojean and G.~Servant,
%``Cosmological expansion in the presence of extra dimensions,''
Phys.\ Rev.\ Lett.\  {\bf 83}, 4245 (1999)
[arXiv:hep-ph/9906523].

\bibitem{BDEL}
P.~Bin\'etruy, C.~Deffayet, U.~Ellwanger and D.~Langlois,
%``Brane cosmological evolution in a bulk with cosmological constant,''
Phys.\ Lett.\ B {\bf 477} (2000) 285
[arXiv:hep-th/9910219].

\bibitem{SMS}
T.~Shiromizu, K.~Maeda and M.~Sasaki,
%``The Einstein equations on the 3-brane world,''
Phys.\ Rev.\ D {\bf 62} (2000) 024012 
[arXiv:gr-qc/9910076].


%%%%%%%%%%%%%%%%%%%%%%%%%%%%%%%%%%%%%%%%%


\bibitem{muko2} 
S.~Mukohyama,
%``Gauge-invariant gravitational perturbations of maximally 
%symmetric  spacetimes,''
Phys.\ Rev.\ D {\bf 62}, 084015 (2000)
[arXiv:hep-th/0004067].


\bibitem{Kodama}
H.~Kodama, A.~Ishibashi and O.~Seto,
%``Brane world cosmology: Gauge-invariant formalism for perturbation,''
Phys.\ Rev.\ D {\bf 62}, 064022 (2000)
[arXiv:hep-th/0004160].

\bibitem{muko1} 
S.~Mukohyama,
%``Perturbation of junction condition and doubly gauge-invariant  variables,''
Class.\ Quant.\ Grav.\  {\bf 17} (2000) 4777
[arXiv:hep-th/0006146].

\bibitem{muko3} 
S.~Mukohyama,
%``Integro-differential equation for brane-world cosmological  perturbations,''
Phys.\ Rev.\ D {\bf 64}, 064006 (2001)
[arXiv:hep-th/0104185].


%%%%%%%%%%%%%%%%%%%%%%%%%%%%%%%%%%%%%%%%%%%%%%%%%%%%%%%%%%%%%

\bibitem{covariant}
R.~Maartens,
%``Cosmological dynamics on the brane,''
Phys.\ Rev.\ D {\bf 62} (2000) 084023
[arXiv:hep-th/0004166].

\bibitem{Chris}
C.~Gordon and R.~Maartens,
%``Density perturbations in the brane world,''
Phys.\ Rev.\ D {\bf 63}, 044022 (2001)
[arXiv:hep-th/0009010].

\bibitem{Leong}
B.~Leong, P.~Dunsby, A.~Challinor and A.~Lasenby,
%``1+3 covariant dynamics of scalar perturbations in braneworlds,''
Phys.\ Rev.\ D {\bf 65}, 104012 (2002)
[arXiv:gr-qc/0111033].


%%%%%%%%%%%%%%%%%%%%%%%%%%%%%%%%%%%%%%%%%%%%%%%%%%%%%%%%%%%%%%%%%

\bibitem{5dpert}
D.~Langlois,
%``Brane cosmological perturbations,''
Phys.\ Rev.\ D {\bf 62} (2000) 126012,
[arXiv:hep-th/0005025].


\bibitem{vandeBruck}
C.~van de Bruck, M.~Dorca, R.~H.~Brandenberger and A.~Lukas,
%``Cosmological perturbations in brane-world theories: Formalism,''
Phys.\ Rev.\ D {\bf 62} (2000) 123515
[arXiv:hep-th/0005032].


\bibitem{Koyama}
K.~Koyama and J.~Soda,
%``Evolution of cosmological perturbations in the brane world,''
Phys.\ Rev.\ D {\bf 62} (2000) 123502
[arXiv:hep-th/0005239].

\bibitem{lmw00}
D.~Langlois, R.~Maartens and D.~Wands,
%``Gravitational waves from inflation on the brane,''
Phys.\ Lett.\ B {\bf 489}, 259 (2000)
[arXiv:hep-th/0006007].

\bibitem{david_prl}
D.~Langlois,
%``Evolution of cosmological perturbations in a brane-universe,''
Phys.\ Rev.\ Lett.\  {\bf 86}, 2212 (2001)
[arXiv:hep-th/0010063].

\bibitem{BMW1}
H.~A.~Bridgman, K.~A.~Malik and D.~Wands,
%``Cosmic vorticity on the brane,''
Phys.\ Rev.\ D {\bf 63}, 084012 (2001)
[arXiv:hep-th/0010133].

\bibitem{neronov}
A.~Neronov and I.~Sachs,
%``On metric perturbations in brane-world scenarios,''
Phys.\ Lett.\ B {\bf 513}, 173 (2001)
[arXiv:hep-th/0011254].

\bibitem{lmsw}
D.~Langlois, R.~Maartens, M.~Sasaki and D.~Wands,
%``Large-scale cosmological perturbations on the brane,''
Phys.\ Rev.\ D {\bf 63}, 084009 (2001)
[arXiv:hep-th/0012044].

\bibitem{deruelle}
N.~Deruelle, T.~Dolezel and J.~Katz,
%``Perturbations of brane worlds,''
Phys.\ Rev.\ D {\bf 63} (2001) 083513
[arXiv:hep-th/0010215].

\bibitem{BMW2} 
H.~A.~Bridgman, K.~A.~Malik and D.~Wands,
%``Cosmological perturbations in the bulk and on the brane,''
Phys.~Rev.~D {\bf 65}, 043502 (2002) 
[arXiv:astro-ph/0107245].

\bibitem{Koyama2}
K.~Koyama and J.~Soda,
%``Bulk gravitational field and cosmological perturbations on the brane,''
Phys.\ Rev.\ D {\bf 65}, 023514 (2002)
[arXiv:hep-th/0108003].

\bibitem{grs01}
D.~S.~Gorbunov, V.~A.~Rubakov and S.~M.~Sibiryakov,
%``Gravity waves from inflating brane or mirrors moving in adS(5),''
JHEP {\bf 0110}, 015 (2001)
[arXiv:hep-th/0108017].

\bibitem{cedric}
C.~Deffayet,
%``On brane world cosmological perturbations,''
Phys.\ Rev.\ D {\bf 66} (2002) 103504 
[arXiv:hep-th/0205084].

\bibitem{riazuelo}
A.~Riazuelo, F.~Vernizzi, D.~Steer and R.~Durrer,
%``Gauge invariant cosmological perturbation theory for braneworlds,''
[arXiv:hep-th/0205220].

\bibitem{Giudice}
G.~F.~Giudice, E.~W.~Kolb, J.~Lesgourgues and A.~Riotto,
%``Transdimensional physics and inflation,''
Phys.\ Rev.\ D {\bf 66}, 083512 (2002)
[arXiv:hep-ph/0207145].

\bibitem{Bardeen}
J.~M.~Bardeen,
%``Gauge Invariant Cosmological Perturbations,''
Phys.\ Rev.\ D {\bf 22},  1882 (1980).

\bibitem{KS}
H.~Kodama and M.~Sasaki,
%``Cosmological Perturbation Theory,''
Prog.\ Theor.\ Phys.\ Suppl.\  {\bf 78} (1984) 1.

\bibitem{MFB}
V.~F.~Mukhanov, H.~A.~Feldman and R.~H.~Brandenberger,
%``Theory of cosmological perturbations. Part 1. Classical perturbations.
% Part 2. Quantum theory of perturbations. Part 3. Extensions,''
Phys.\ Rept.\  {\bf 215} (1992) 203.

\bibitem{reviews}
R.~Maartens,
%``Geometry and dynamics of the brane-world,''
[arXiv:gr-qc/0101059]; 
D.~Langlois,
%``Brane cosmology: An introduction,''
Prog.\ Theor.\ Phys.\ Suppl.\  {\bf 148} (2003) 181
[arXiv:hep-th/0209261]; 
N.~Deruelle, [arXiv:gr-qc/0301036];
P.~Brax and C.~van de Bruck,
%``Cosmology and brane worlds: A review,''
[arXiv:hep-th/0303095].

\bibitem{stewart} 
J.~M.~Stewart, Class.~Quantum Grav.{\bf{7}} 
1169 (1990).

\bibitem{wald}
R. M. Wald, 
{\it General Relativity}, (Univ. Chicago Press,
Chicago, 1984).

\bibitem{israel}
W.~Israel,
%``Singular Hypersurfaces And Thin Shells In General Relativity,''
Nuovo Cim.\ B {\bf 44S10} (1966) 1 [Erratum-ibid.\ B {\bf 48}
(1966) 463].

\bibitem{4Dmatching}
N.~Deruelle and V.~F.~Mukhanov,
%``On matching conditions for cosmological perturbations,''
Phys.\ Rev.\ D {\bf 52} (1995) 5549
[arXiv:gr-qc/9503050].

\bibitem{bounce} 
C.~Gordon and N.~Turok,
%``Cosmological perturbations through a general relativistic bounce,''
[arXiv:hep-th/0206138].




%%%%%%%%%%%%%%%%%%%%%%%%%%
\end{thebibliography}
\end{document}